\documentclass[USenglish]{article}

\usepackage[utf8]{inputenc}

\usepackage{color}
\usepackage{amssymb,amsmath,latexsym}
\usepackage{graphicx}
\usepackage{url}
\usepackage{natbib}
\usepackage{hyperref}
\newcommand{\footremember}[2]{%
    \footnote{#2}
    \newcounter{#1}
    \setcounter{#1}{\value{footnote}}%
}

\defcitealias{Jumpstart2012}{Jumpstart Group, 2012}

\begin{document}

\title{Machine learning for metagenomics: methods and tools}

\author{
Hayssam Soueidan \footremember{cbib}{ Bordeaux Bioinformatics Center, Universit\'e de Bordeaux, France}%
  \and Macha Nikolski \footremember{labri}{Corresponding author: \texttt{macha.nikolski@labri.fr}; Bordeaux Bioinformatics Center, Universit\'e de Bordeaux, France and LaBRI, Universit\'e de Bordeaux, France}%
  }
\date{}
   
\date{\today}

\maketitle

\begin{abstract}
 Owing to the complexity and variability of metagenomic studies, modern machine learning approaches have seen increased usage to answer a variety of question encompassing the full range of metagenomic NGS data analysis. We review here the contribution of machine learning techniques for the field of metagenomics, by presenting known successful approaches in a unified framework. This review focuses on five important metagenomic problems: OTU-clustering, binning, taxonomic profling and assignment, comparative metagenomics and gene prediction. For each of these problems, we identify the most prominent methods, summarize the machine learning approaches used and put them into perspective of similar methods. We conclude our review looking further ahead at the challenge posed by the analysis of interactions within microbial communities and different environments, in a field one could call ``integrative metagenomics''.
\end{abstract}

\section{Introduction}

While genomics is the research field relative to the study of the genome of any organism, metagenomics is the term for the research that focuses on many genomes at the same time, as typical in some sections of environmental study. Metagenomics recognizes the need to develop computational methods that enable understanding the genetic composition and activities of communities of species so complex that they can only be sampled, never completely characterized. 

Analysis of microbial communities has been until recently a complicated task due to their high diversity and the fact that many of these organisms can not be cultivated. Metagenomics has essentially emerged as  the only way to characterize these unculturable communities. Though ``traditional'' genomic studies are challenging, metagenomic studies push the requirement for highly scalable computational solutions further. Indeed, computational analysis is even more important for metagenomic analysis due not only to the large amount of metagenomic data, but also to the new questions introduced by metagenomic projects such as simultaneous assembly of multiple genomes, community-realized functions or host-microbe interactions \citep{Metagenomics2}.

Metagenomics has been popularized by the studies of bacterial communities and is still largely dominated by such applications. Indeed, the word itself is often employed in this sense. However, such a reductive view does not grant enough credit to the lively research in other areas such as viral or fungal metagenomics, and large-scale environmental studies that generate extremely heterogeneous data in terms of their origin.

Bazinet and Cummings have listed in their review 25 tools \citep{Bazinet2012} and this collection continuously expands. The specific challenges of "Big Data metagenomics" that the field is currently facing promote computational solutions originating from data science. In particular, recent years have brought a large set of work in machine learning applied to metagenomics. Indeed, machine learning is the methodology  of finding patterns and making predictions from data, based on multivariate statistics, data mining and pattern recognition. 

Machine learning currently offers some of the most promising tools for building predictive models for classification of biological data (see e.g. \cite{mlbio2015}). Various biological applications cover the entire spectrum of machine learning problems including supervised learning, unsupervised learning (or clustering), and model construction. Moreover, most of biological data, such as those produced in metagenomic studies, are both unbalanced and heterogeneous, thus meeting the current challenges of machine learning in the era of Big Data. 

The goal of this review is to examine the contribution of machine learning techniques for metagenomics, that is, answer the question \emph{to what extent does machine learning contribute to the study of microbial communities and environmental samples?} We will first briefly introduce the scientific fundamentals of machine learning. In the following sections we will illustrate how these techniques are helpful in answering questions of metagenomic data analysis. We will describe a certain number of methods and tools to this end, though we will not cover them exhaustively. Finally, we will speculate on the possible future directions of this research.

\section{Machine Learning Fundamentals}


\paragraph*{General principles}
The defining characteristic of machine learning (ML) systems is that they can improve with experience. Contrary to a system employing a classical deterministic algorithm, an ML system will exploit patterns observed in datasets to fine-tune and adapt its decisions. As defined by \citet{Mitchell1997}, a typical ML program requires at least three different components: 
\begin{enumerate}
\item \emph{experience}, in the form of data,
\item \emph{task}, in the form of an output of the algorithm, 
\item \emph{objective}, in the form of performance measurement of a given output. 
\end{enumerate}
A program is said to ``learn'' if its performance for a given task improves with experience. 

To design a program able to learn how to solve a given task, Flach \citep{FLACH2012} indicates the following requirements. First, data from the domain must be mapped to {\bf features} constituting an (often finite) representation given as input to a {\bf model} that will map it to an output. To build this model, some example data -- often called {\bf training data} -- are fed into a {\bf learning algorithm} that will search and try to identify an optimal model (w.r.t. the objective function) from a {\bf hypothesis space}. 

Learning algorithms are often grouped by how the model that they learn is represented in this hypothesis space.  Such spaces and potential hypotheses are usually structured and reduced in order to obtain a "well-behaved" hypothesis space where the learning problem can be redefined mathematically in terms of a numerical optimization problem. 

In the field of metagenomics, the most common questions that we identified as being amenable to learning are: OTU clustering, binning, taxonomic assignment, gene prediction and comparative metagenomics. These questions involve three types of machine learning tasks (see Table \ref{fig:examples-ml}): 
\begin{enumerate}
\item  \emph{classification}, where the task output is a discrete variable,
\item  \emph{clustering}, where the task output is a (possibly soft) partition of input data,
\item  \emph{dimensionality reduction}, where the task output is a lower-dimensional representation of the input data. 
\end{enumerate}

Another useful distinction is between supervised and unsupervised problems. A problem is said to be {\bf supervised} if the output variable is provided in the training data. For example, distinguishing metagenomic samples based on known phenotypical traits is a supervised problem. The goal of supervised methods is thus to build a model from a set of labeled data points that can predict the correct category of unlabeled future data. Labels can be provided by any type of important metadata, such as the species of the microbe or of the host, but they have to be discrete and scalar. The possibility to classify unlabeled data is especially useful when alternative methods for obtaining data labels are difficult such as it is the case, for example, for unculturable bacteria. On the other hand, {\bf unsupervised} methods assume no labeled response but rather seek to determine hidden structure in the data. The most common application of unsupervised methods are for cluster analysis as well as dimensionality reduction. 

\paragraph*{Components of ML systems}

As indicated by \citet{Domingos2012}, any learning algorithm can be broken down into three components: (i) {\bf representation}: the structure of the hypothesis space, (ii) {\bf evaluation}: how a model will be evaluated, and (iii) {\bf optimization}: how to navigate in the hypothesis space. 

Some examples of hypothesis spaces / learning algorithm that are relevant for metagenomics are shown in Table \ref{fig:examples-ml}.
\begin{center}
\begin{table}[htbp]
\includegraphics[width=\textwidth]{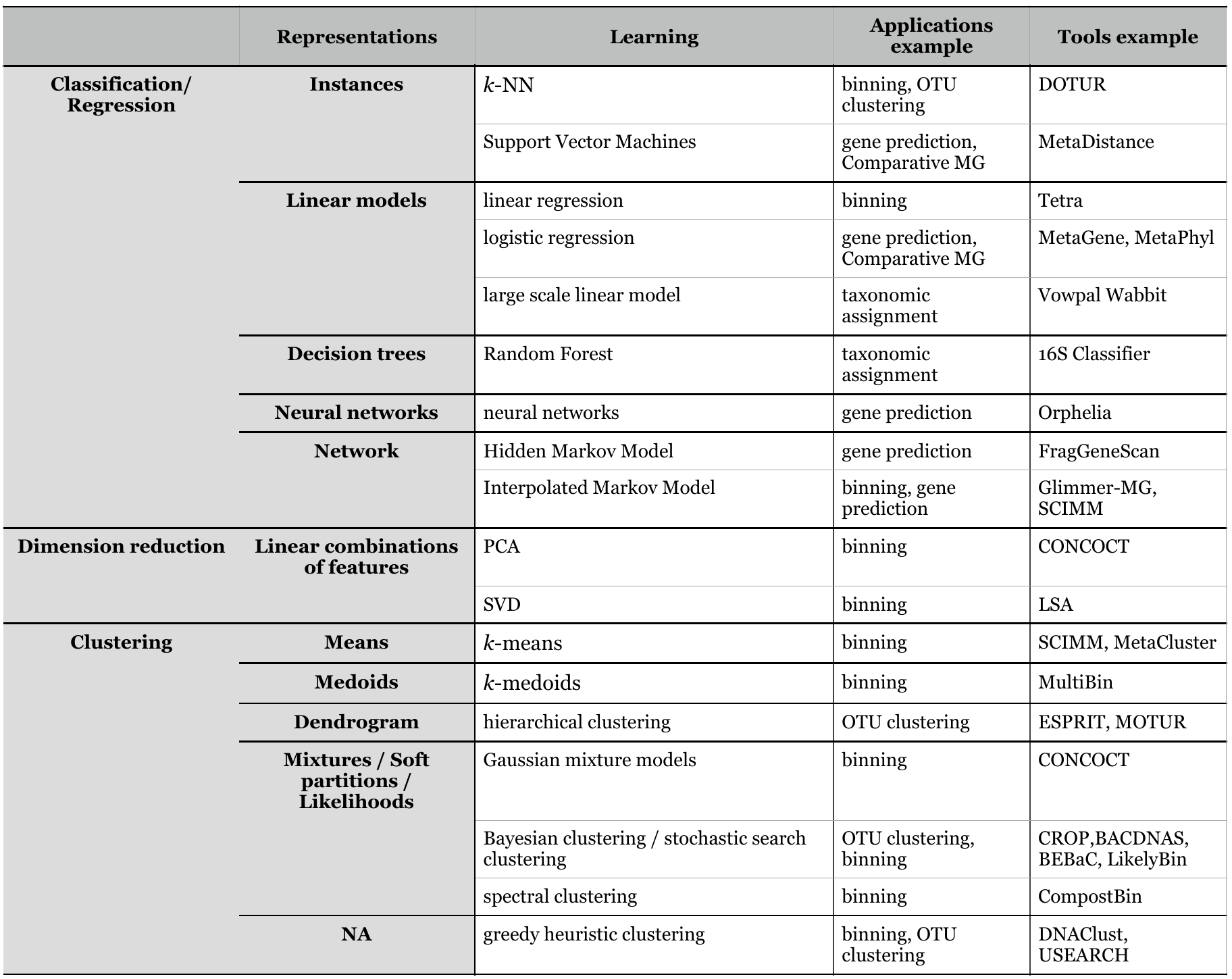}
\caption{\label{fig:examples-ml}Examples of machine learning approaches for metagenomic applications. Methods are classified according to the three identified types of machine learning tasks: {\bf Classification}, {\bf Dimension reduction} and {\bf Clustering}. These tasks can be further analysed by the type of representation of the hypothesis space (column {\bf Representations}) and the learning  algorithm that is applied to training data (column {\bf Learning}). Finally, we mention types of metagenomic analyses to which these approaches have been applied (column {\bf Application example}) and show examples of existing software that implements them (column {\bf Tools example}).}
\end{table}
\end{center}

In the case of supervised learning, the performance of a model shouldn't be measured on the training data. The main reason is that the real objective of inductive learning is not to perform perfectly on data that was already seen, but to generalize well for previously unseen data. The phenomenon of poor performance on unseen data is called ``overfitting'', and a simple way to ensure its absence is to leave out part of the data for evaluation purposes. It is however worth noting that different representations suffer differently from overfitting. Usually, the more complex the hypotheses are, the more care should be put in overfitting avoidance. For example, Na{\"i}ve Bayes models have a simple representation (basically, just conditional probabilities / counts) and their simplicity makes them less prone to overfitting than more complex models (e.g. Random Forests), but their simplicity doesn't allow complex decision functions to be represented. On the other hand, very complex models (such as e.g. neural networks) can approximate universal decision functions with any level of precision, but require tremendous training data and usually can not be trained to optimality. 

\paragraph*{Features}
Besides the learning algorithm and the model \emph{per se}, the most important component of a learning system is how features are extracted from the domain data, a process known as {\bf feature engineering}. As stated by Flach \citep{FLACH2012}, ``features are the workhorse of machine learning''. In our case, reads originating from an NGS project must be mapped to a domain that can be processed by the learning algorithm. Often, this domain is a vector in a multidimensional space. The combination of (i) how to map NGS reads into vector representation, (ii) learning algorithm and (iii) how a metagenomic question is formulated as a task is actually what distinguishes most of the methods described in this review. 

A very natural and simple representation of DNA sequences as a numerical vector is the $k$-mer frequencies representation. For a given integer $k$, count the number of times any nucleotide subsequence of length $k$ appears in the read. Subsequent transformation can enrich this representation (e.g. robust estimations, variance stabilization, normalization etc.). Naturally, the precision of this representation increases with $k$, but this comes with the cost of much higher computational and memory requirements needed to represent each read. For example, even for values of $k$ as small as $k=12$, each read will be represented by $4^{12} \approx 16 \times 10^6$ dimensional vector, with a memory cost that is out of reach of today's computers, given a reasonably-sized metagenomics project. In fact, a more pervasive issue appears as $k$ grows: generalizing a decision function becomes exponentially harder as we increase the dimensionality of the space, since only a very limited fraction of the feature space can be observed in the training data. Pushing this argument to its extreme, it is straightforward to observe that most $21$-mers are actually unique to species present in public sequence databases \citep{kurtz2008new}. On one hand, this implies that we can recognize any known species with high precision, but on the other hand, this implies that novel species can not be classified and thus that such a system has close to zero generalization power. This is indeed the case for most large $k$-mer indexing schemes (e.g., LMAT \citep{Ames2013}). Even for moderate values of $k$, it might still be the case that the feature space is too large given the training data, but that lower values of $k$ yield too low resolution. In this case, it is possible to learn a lower-dimension manifold in which the data is projected, and instances of such techniques (e.g. PCA, SOM) are used by several of the tools reviewed here. 

\section{Metagenomics: data and questions}

Modern metagenomic analysis involves a variety of options: genetic diversity profiling of samples, identification of present taxa (taxonomic assignment), assembly of sequencing reads, clustering and binning, prediction of protein-coding genes and further functional analysis of community-realized functions, without forgetting the comparative aspects of multi-sample studies. Machine learning has penetrated in all these applications, even assembly, which is traditionally solved by data-structure and algorithmic approaches, has been tackled by machine learning \citep{Afiahayati2015}. In this review we will focus on five important questions of metagenomic  studies from the perspective of machine learning: OTU clustering, binning, taxonomic assignment, comparative metagenomics and gene prediction.

\paragraph*{Terminology} The abundance of literature on metagenomic analysis has generated quite a variable terminology, which makes navigating the underlying computational concepts a complex task. For the sake of consistency, we will keep the same terminology throughout this paper. In particular, we will use the term of \emph{assignment} to mean the assignment of a given sequence to a specific taxonomic group (see section \ref{sec:tax}). We will thoroughly distinguish this from both \emph{OTU clustering} and \emph{binning} which correspond to the same machine learning problem of clustering as both refer to the grouping of a data set into subgroups which are distinct from one another. Being unsupervised problems by nature, these subgroups may remain unlabeled. However, we will distinguish them by using the term \emph{OTU clustering} exclusively for clustering of marker gene sequencing data (see section \ref{sec:otu}), while \emph{binning} will be used for the case of whole metagenome untargeted data (see section \ref{sec:bin}).

Notice that all these different analyses perform a \emph{classification}, the difference being that in one case it is unsupervised (OTU clustering and binning) and in another, supervised (assignment). A lot (but not all) of the methods presented below contain both some form of grouping and an assignment step. We have decided to discuss them in the sections that correspond best to the computationally innovative part of such papers.

\subsection*{Data}

Metagenomic studies differ by a number of elements. First and foremost is the kind of data that is collected, which largely determines the downstream analysis. 

\paragraph*{Amplicon sequencing} In many metagenomic studies only partial genomic information is used.  DNA is extracted from all the cells of the the sample (e.g., sea, soil, gut), a set of genomic markers is chosen and then targeted and amplified by PCR.  Consequently, only marker genes are sequenced. The most widely used marker gene for amplicon sequencing is the small subunit ribosomal RNA (16S) locus, that is both a taxonomically and phylogenetically informative marker for both bacteria and archaea \citep{Pace1986, Hugenholtz1996}. Other markers such as the small subunit 18S, the large subunit (LSU: 23S/25S/28S) and the  highly variable spacers ITS1 and ITS2 of the internal transcribed spacer (ITS) region have also been used, in particular for fungi, higher plants, insects and microalgae \citep{Lindahl2013, Koetschan2010, Cuadros2013}. 

Such targeted sequencing simplifies the analysis for two reasons. First, the amount of data remains reasonable (for a high-throughput analysis) and second, known marker genes' taxonomic classification is available -- in particular for bacterial communities  -- through reference taxonomies such as, for example, in RDP \citep{cole.ea:2009} or  Greengenes \citep{mcdonald.ea:2012} databases. 

Many synonyms for amplicon studies can be found in literature, the most frequent being \textbf{metabarcoding} and \textbf{targeted metagenomics}.

\paragraph*{Untargeted sequencing} Whole metagenome sequencing without targeted amplification provides the possibility to tap into unknown species, but generates a substantial increase in the volume of data. Here, one of the important aspects is the representativity of the extracted DNA with respect to taxa that compose the sample, and various protocols exist \citep{Delmont2011,Burke2009}. If the community of interest is associated with a host than its DNA has to be separated by either fractionation or selective lysis to ensure that minimal host DNA is present. Physical filtering (fractionation) can also be used when only part of the community is of interest to the analysis. The organisms are then separated by size, leading to untargeted sequencing of filtered samples. Various physical and molecular extraction / filtering techniques exist and can be used to separate eukaryotic, prokaryotic, or viral particles. In particular, size-based filtering is widely used for separating viruses prior to studying the composition of viral communities \citep{John2011, Hurwitz2012}.

\paragraph*{Choice of data}
Notice that the choice of data to be used in a given metagenomic study is not neutral since it determines the final results of the analysis. For example, in the case of amplicon sequencing, the choice of amplicons is known to influence the resulting clusters. Regions of marker genes with high inter-species and low intra-species variability are typically chosen to be sequenced (e.g. V1 through V9 for 16S). They are flanked by conserved sequences that enable PCR amplification of targeted sequences (amplicons) using universal or specifically designed primers. However, depending on the chosen variable regions and primers, the resulting OTUs clusters and estimations of diversity may not be the same (see for example \citepalias{Jumpstart2012}, \citep{Bella2013}). Moreover, amplicon sequencing excludes unknown species or highly diverged microbes from the analysis. Analyzing whole-metagenome untargeted sequencing data is the most computationally challenging task in metagenomics for the reasons of both data volume and variety. For example, environmental samples can contain genetic material from hundreds to tens of thousands of species of different abundances, some estimations placing the number of species for certain environments in the range of millions \citep{diversity}. Moreover, a large fraction species collected this way may not be known, potentially leading to annotation biases since the analysis relies on the genomes present in the current databases. In this respect, identification of viruses remains especially challenging \citep{Soueidan2014}. Sequencing filtered samples alleviates the computational burden downstream, thus allowing efficient contig assembly \citep{Coetzee2010, Minot2012}. However, it introduces certain biases, such as artificially including or excluding certain species.

\section{OTU clustering}
\label{sec:otu}
A wide-spread approach to explore the genetic diversity of a microbial community is to analyse amplicon sequencing data (or barcode sequences) by grouping it into Operational Taxonomic Units (OTUs). Historically, ribosomal RNAs are the most used basis for the reconstruction of microbial phylogeny \citep{ludwig1994bacterial} and the definition of the corresponding taxa. Thus, rRNA approach has quite naturally become the most popular one for defining taxa in targeted bacterial studies \citep{blaxter2005defining} and has been since generalized to other marker genes. The OTUs are defined at varying levels of identity (often 97\%,  see e.g.\citep{koeppel2013surprisingly}), and can then be used to estimate the diversity and composition in terms of taxa of a given sample. Methods for OTU clustering typically require a distance function $d$ and a certain threshold $t$ under which two sequences are considered to be sufficiently close. One of the characteristics of OTU-based approaches is that all sequences are clustered into OTUs, including both species for which annotation is absent in the public databases as well as novel species. Moreover, OTU clustering has the double advantage of being a light-weight approach in terms of input data (amplicon sequencing) and consequent computation, as well as of producing high-quality groups precisely because amplicon sequences are by definition taxa-specific and different between species. Such data are an ideal setting for highly sensitive and specific machine learning methods. We recapitulate OTU clustering methods presented in this review in Table \ref{fig:otu-tools}.

While quite computationally performant, OTU-based methods also possess certain intrinsic limitations. Most importantly, the analysis and biologically meaningful interpretation of the resulting OTUs can be problematic, especially in the case of unknown species. Also, while the threshold $t$ greatly influences the inferred clusters, it is difficult to choose \emph{a priori} given that the evolutionary rates for marker genes can vary greatly \citep{patwardhan2014molecular}. Furthermore, due to sequencing errors, wrong choice of $t$ can result in an artificial inflation of OTUs \citep{kunin2010wrinkles}. Moreover, a number of studies point out that different clustering methods  provide non-equivalent clusters for the same dataset, quantitatively in terms of total number of clusters and  OTU size distributions as well as qualitatively in terms of cluster composition \citep{Huse2010, Schloss2011, Sun2011, Bonder2012}. This is particularly salient in the presence of sequencing errors.
\begin{center}
\begin{table}[htbp]
\includegraphics[width=\textwidth]{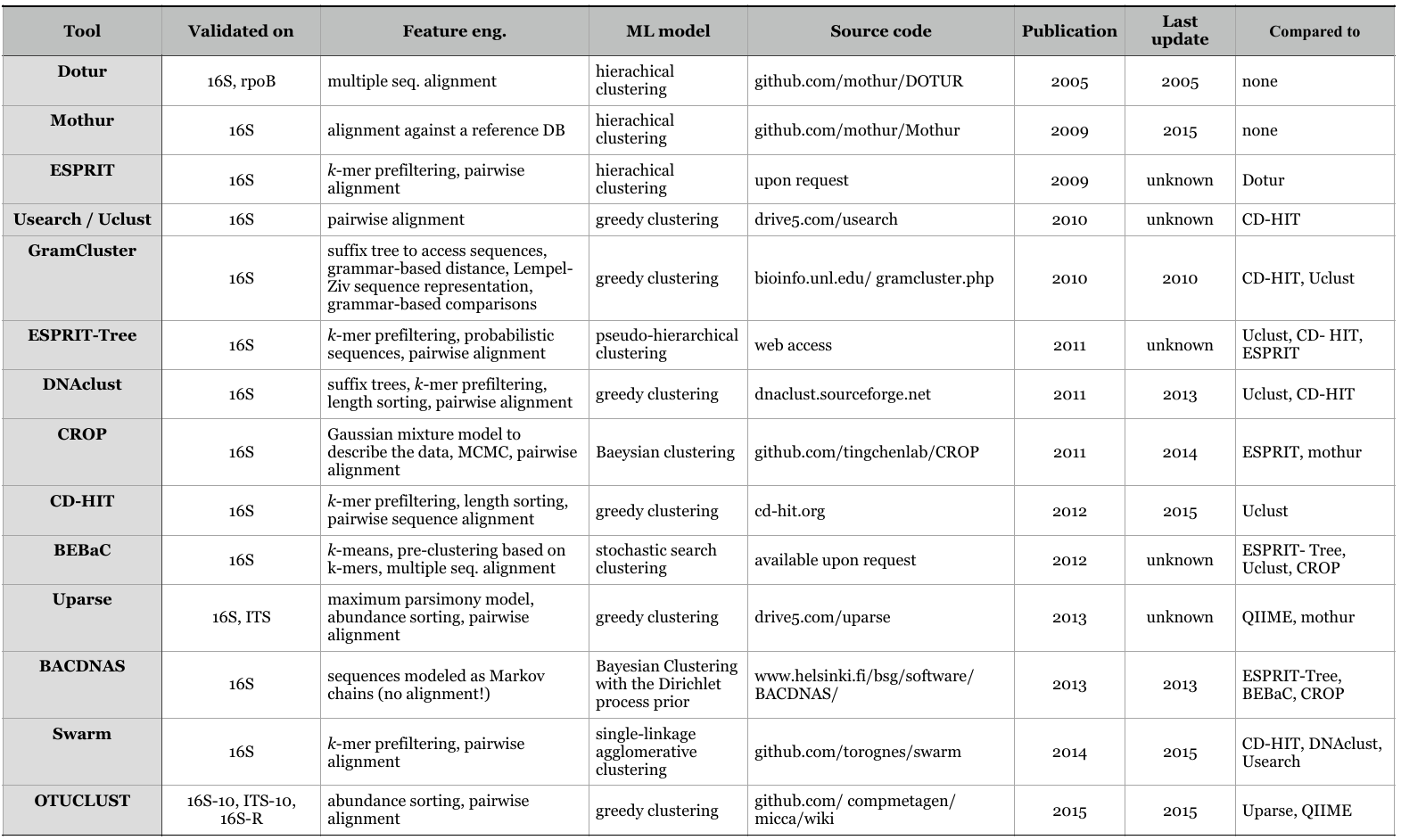}
\caption{\label{fig:otu-tools}Characteristics of tools for OTU clustering. Shown are types of data on which the method has been {\bf validated}, how features have been extracted (column {\bf Feature eng.}), the machine learning algorithm (column {\bf ML method}), link to the {\bf Source code}, year of {\bf publication} as well as the last source code update (column {\bf Last update}), and finally tools against which the performance has been evaluated (column {\bf Compared to}).}
\end{table}
\end{center}

\subsection{Hierarchical clustering}
\label{sec:hierarchical-OTU}
Hierarchical clustering algorithms such as the nearest neighbor (single linkage clustering), the furthest neighbor (complete linkage clustering), the weighted neighbor, and the average neighbor (unweighted-pair group method using average linkages, UPGMA) are widely used  for OTU clustering. These algorithms require the computation of the complete pairwise distance matrix between sequences. Based on this matrix,  a hierarchical tree is constructed. Given a user-specified threshold $t$, leaf nodes within the cutoff are assigned to an OTU. The main disadvantage of the hierarchical clustering strategy is the overestimation of the number of OTUs, especially in the presence of sequencing errors, resulting in skewed OTU abundances. 

Alignment-based methods suffer from the high computational cost and methods such as {\bf Dotur} \citep{Schloss2005} (based on multiple sequence alignment) and {\bf Mothur} \citep{Schloss2009} (based on Needleman-Wunsch alignments  against a pre-aligned reference database) do not scale up very well despite the use of sparse matrices that represent the unique sequences only in the case of Mothur.

This gave rise to methods that attempt to achieve better scalability, such as {\bf ESPRIT} \citep{Sun2009} that implements a complete-linkage hierarchical clustering and minimizes the memory usage by using an on-line approach, and {\bf ESPRIT-Tree} \citep{Cai2011} that uses a probabilistic distance and a set of heuristics in order to avoid all pairwise distances. The computational speedup is also achieved by avoiding to align pairs of sequences for which it is easy to deduce from their $k$-mer comparison that the alignment will not be good enough.

A recent tool called {\bf Swarm} \citep{Mahe2014} is an agglomerative, unsupervised, single-linkage clustering algorithm that avoids the use of a global threshold $t$. First, it attempts to iteratively cluster highly similar amplicons using a local user-defined threshold and then refines the results by using clusters internal structure and amplicon abundances.

\subsection{Heuristic clustering} 
\label{sec:greedy-OTU}

Amplicon sequencing projects, especially in the case of environmental studies, produce constantly increasing datasets. For this reason, in order to compute OTUs, two important requirements are speed and scalability. However, exact clustering algorithms described in section \ref{sec:hierarchical-OTU} do not scale  well, requiring at worst the square of the number of input sequences comparisons. This has given rise to methods based on \emph{heuristic greedy clustering}. Given a set of sequences $S$, a sequence $s \in S$ is chosen and it becomes the centroid of a new OTU cluster. This sequence $s$ is then compared to all remaining sequences $S \setminus s$. Given a distance function $d$ and a threshold $t$, all sequences $s' \in S$ such that $d(s,s') < t$ are added to the OTU and removed from $S$. Since each sequence is considered only once, heuristic greedy algorithms are usually fast. Notice, that the greedy approach provides results that are dependent on the order of input sequences.

A lot of methods  in this category sort input sequences by length or abundance, such as for example  {\bf Uclust} \citep{Edgar2010}, {\bf DNAclust} \citep{Ghodsi2011} and {\bf CD-HIT} \citep{suzek2007uniref,Fu2012} and try to avoid costly all-against-all sequence alignment by a pre-filtering $k$-mer based step. Moreover, {\bf GramCluster} \citep{Russell2010} and DNAclust \citep{Ghodsi2011} index the input dataset by a suffix tree for efficiency.

While remaining in the scope of greedy clustering, the {\bf Uparse} \citep{Edgar2013} and {\bf OTUCLUST} of {\bf MICCA} \citep{Albanese2015} emphasize the need to rely on high quality sequences only and  both include a number of steps for quality filtering, trimming and chimera filtering.

\subsection{Model-based clustering}
\label{sec:model-OTU}

Model-based clustering aims to circumvent the overestimation of OTUs that due to the limitations of choosing an \emph{a priori} threshold $t$. 

For example, {\bf CROP} \citep{Hao2011} uses an unsupervised Bayesian clustering approach. The model relies on the notion of probability that a given sequence $s$ belongs to a cluster. This probability is defined as function of the distance between the sequence $s$ and the sequence that is in the center of the cluster. Moreover, CROP applies the divide-and-conquer principle by dividing the dataset into small subsets and performing Bayesian clustering on the subsets. Thus generated clusters are replaced by their consensus sequences on which a final step of Bayesian clustering is performed  in order to obtain the OTUs. 

In \citep{Cheng2012}  the authors introduce a probabilistic model-based {\bf BEBaC} method. It is based on the calculation of an unnormalized posterior probability for an arbitrary partition of the reads, such that any two partitions can be compared. The best partition is then determined by a stochastic search process over the partition space.

In \citep{Jaaskinen2013} the authors introduce the {\bf BACDNAS} method that models sequences by Markov chains of fixed order and an expectation-maximization algorithm is used for learning a partition of these Markov chains. The prior distribution for partitions is the Dirichlet process prior.

\section{Binning}
\label{sec:bin}
The goal of \emph{binning} is to establish the taxonomic profile of a given sample and it is often used in the case of environmental studies. We differentiate binning from OTU clustering mainly in terms of input data: while OTU clustering is used in targeted studies, binning deals with reads coming from any genomic region of any genome present in the sample. Binning is thus the most suited approach for studying complex communities. Nevertheless, almost all available methods are developed for bacterial communities, with the notable exception of MetaVir \citep{Roux2013} that aims at the analysis of viromes. Analysis of other communities, such as for example fungi, are often done by ad-hoc methods or by hijacking software tools designed for bacteria (see e.g. \citep{Lindahl2013, Cuadros2013}).

In terms of computation, binning consists of assigning each read to a group, called \emph{bin}, with the expected property that each bin consists of reads originating from the same taxon. Although some alignment-based methods exist (not detailed in this review), most existing computational tools for binning utilize sequence $k$-mer composition. Indeed, as observed by \citet{kariin1995}, the distribution of $k$-mer composition is stable across a single genome and varies between genomes, even when only dinucleotides (di-mers) are considered. This observation underlies many metagenomic binning approaches.

In the existing literature, machine learning-based classification algorithms used for binning are categorized into supervised, semi-supervised, and unsupervised classes, depending on whether a training set of labeled data is used to build their models or not. We consider that the former case corresponds to solving the \emph{taxonomic assignment} problem, that we describe in the next section. In this section, we only consider unsupervised or semi-supervised methods for the binning problem exclusively, that is methods that group sequences into bins without consideration of external reference sequences or taxonomic information in the same way it is done for OTU clustering. Indeed, it has been observed that in environmental samples, up to 99\% of sampled bacteria are unknown \citep{eisen2007environmental}, hence methods relying on existing datasets will very likely miss important properties of samples. To solve the binning problem, various measures of similarity such as GC content, oligomer frequencies, the abundance of genes or contig coverage  can be used during the inference. See Table \ref{fig:bin-tools} for the summary of binning methods presented in this review.

\begin{center}
\begin{table}[htbp]
\includegraphics[width=\textwidth]{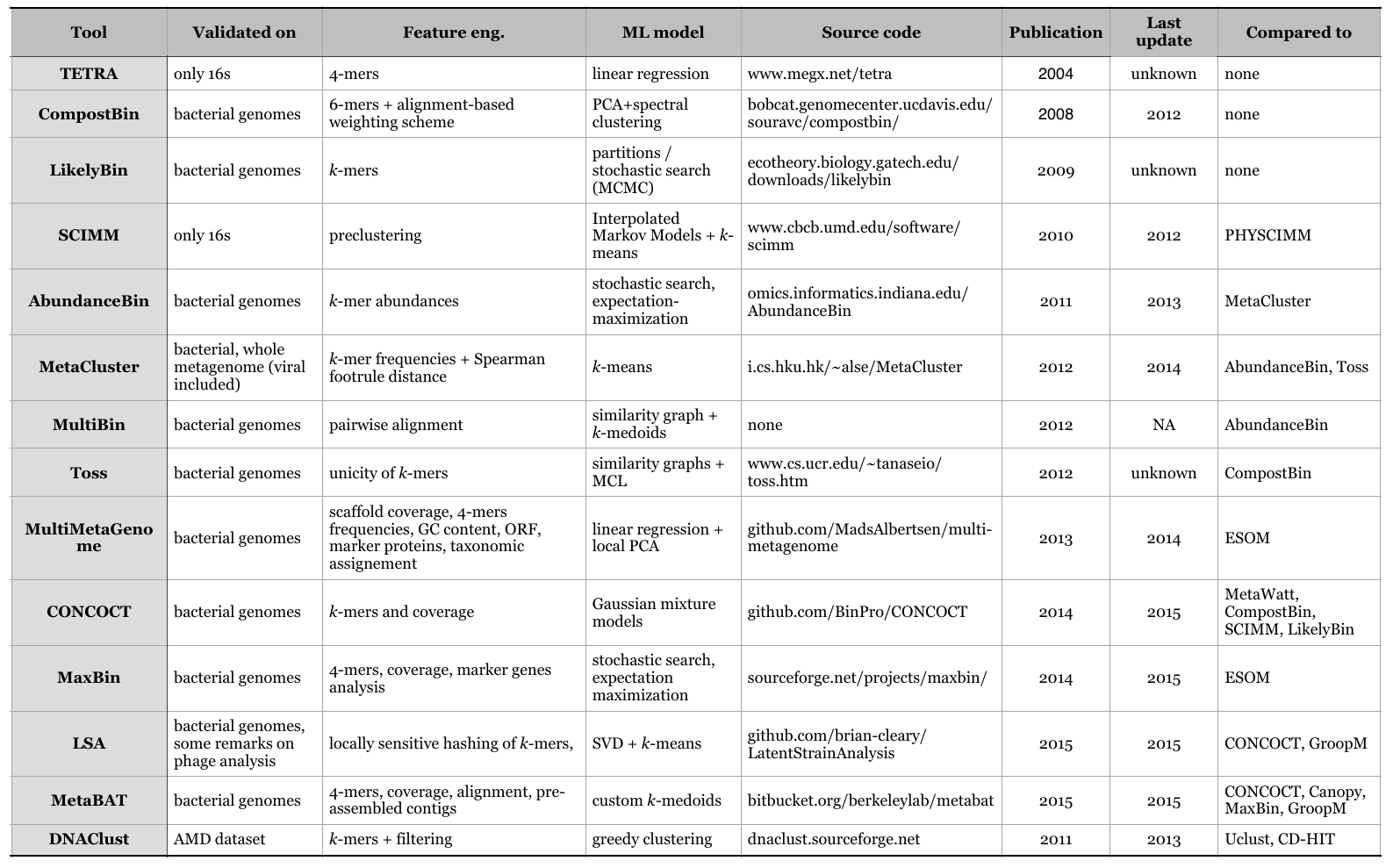}
\caption{\label{fig:bin-tools}Characteristics of binning tools. Shown are types of data on which the method has been {\bf validated}, how features have been extracted (column {\bf Feature eng.}), the machine learning algorithm (column {\bf ML method}), link to the {\bf Source code}, year of {\bf publication} as well as the last source code update (column {\bf Last update}), and finally tools against which the performance has been evaluated (column {\bf Compared to}).}
\end{table}
\end{center}

Binning is a computationally expensive task since the complete untagged environmental datasets are both large and heterogeneous due to the high-complexity of communities in terms of number of species, such as exemplified by ocean microbial communities (see the Science special issue on Tara Oceans project \citep{Tara}) and the human microbiome project (\url{https://commonfund.nih.gov/hmp/index}).

Binning methods are known to perform better on either longer reads or on pre-assembled datasets. It has been shown that even partial assembly improves the strength of the taxonomic signal contained in individual short reads, and that even in the case of increased chimericity \citep{Teeling, Mende}. Consequently, given a metagenomic dataset, a lot of methods suppose that at least partial assembly has been performed to obtain contigs of the size not less than $1000$bp.

\subsection{Fully unsupervised clustering}
\label{secunsupervised-bin}
To the extent of our knowledge, the first application of $k$-mer distribution to the binning problem was proposed by Teeling in 2004 with a tool called {\bf Tetra} \citep{Teeling2004}. Tetra computes a $z$-score for each possible 4-mer then compares two fragments by Pearson correlation over all 4-mers. However, the author indicates that this method works well for very large sequences only (in the range of 40kb). 

More recently, {\bf CompostBin} \citep{Chatterji2008} uses an unsupervised clustering scheme based on graph cuts. First,  principal component analysis extracts meaningful $k$-mer features and then the tool applies a normalized cuts clustering technique (spectral clustering) to group sequences into taxa-specific clusters. {\bf MetaCluster} \citep{Wang2012, Wang2012b} is a collection of unsupervised methods based on $k$-means clustering. The authors make uses of three observations that are the basis of most unsupervised binning strategies. 
\begin{enumerate}
\item $k$-mer frequencies from reads of a genome are linearly proportional to the genome abundance, 
\item long $k$-mers are unique, and
\item short $k$-mers frequency distribution are similar for similar genomes. 
\end{enumerate}
Furthermore, observing that binning is sensitive to the range of abundance (for samples containing both extremely-low and high abundance), the authors propose to separate reads originating from low abundance species from reads originating from high abundance species. The group of high-abundance species' reads are then processed separately using the $k$-means algorithm on the 5-mer representation, while the group of low-abundance species' reads are binned using 4-mer's frequencies. 

In its turn, {\bf Toss} \citep{Tanaseichuk2012} groups all unique $k$-mers into clusters and then merges these clusters based on $k$-mer repeat information by the Markov Cluster algorithm. In the case, where samples are expected to have highly different abundance levels of individual species, Toss relies on AbundanceBin \citep{Wu2011}  to pre-cluster reads.  {\bf MetaVir} \citep{Roux2013} was developed to deal specifically with filtered viral sequencing data and is based on $k$-mers of size 2, 3 and 4; hierarchical clustering is performed based on the Euclidean distance between $k$-mer frequency vectors. 

A recent method called {\bf Latent Strain Analysis}  (LSA) \citep{Cleary2015} tries to leverage modern machine-learning techniques such as feature hashing \citep{weinberger2009feature} and online dimensionality reduction. It is a binning method based on the hyperplane clustering approach and relying on a complex-space hashing approach. They devise a $k$-mer hashing function that has the property of being locally sensitive: similar $k$-mers have a much higher probability of generating a hash collision. Such $k$-mers are placed in the same row within a matrix that recapitulates $k$-mer abundance for each sample. This matrix is then reduced using singular value decomposition (SVD) - meaning that it is performed in the space of $k$-mer co-variation - and reads are then assigned to the resulting $k$-mer clusters. The intuition behind the method is that co-variance information provided by SVD represents the relationships between $k$-mers found in the same genome.

Orthogonally, one of the approaches that has been fairly used in literature is self-organizing maps (SOMs), which groups sequences by oligonucleotide frequencies \citep{Dick2009} and read coverage levels \citep{Sharon2013}. Although vastly used, application of SOMs requires a manual step, where the user selects sets of reads (or scaffolds) that should be binned together.

\subsection{Model-based clustering}
\label{model-bin}

Another family of approaches to solve the binning problem relies on inference of a probabilistic model. These methods are unsupervised, and mostly use expectation maximization to estimate model parameter. 

The main goal of {\bf LikelyBin} \citep{Kislyuk2009}  is to bin together relatively short reads (400nt) from low complexity samples (between 2 and 10 species) that show sufficient evidence of genomic diversity. To this end, an inference scheme based on maximum likelihood principle and Markov Chain Monte Carlo estimates a nucleotide model (a Markov model with order ranging from 3 to 5 ) and then assigns each read to the most likely model. 

The Markov model of LikelyBin has been generalized to varying order Markov chains by \citep{Kelley2010} in a tool called {\bf SCIMM}. Instead of having a fixed order, varying order chains (termed Interpolated Markov Models) interpolate between all models of lower order. Furthermore, the authors propose an expectation maximization scheme similar to the $k$-means algorithm, but where each cluster is modeled as an Interpolated Markov chain and cluster assignment follows maximum likelihood. Cluster initialization (one of the major problems of $k$-means clustering) can be obtained by running  LikelyBin or CompostBin. Interestingly, the authors also suggest a supervised scheme for cluster initialization based on taxonomic assignment obtained by the tool Phymm (described in section \ref{sec:tax}). 

{\bf AbundanceBin} \citep{Wu2011} was the first method to rely on information different from similarity and composition. Indeed, it uses $k$-mer abundance information in order to separate reads from genomes that have different abundance levels.  AbundanceBin first computes frequencies of all $k$-mers in a metagenomic dataset and then, assuming that these frequencies come from a mixture of Poisson distributions, models  the original reads as a mixture of these distributions. An expectation-maximization algorithm is further used to find the distributions' parameters. Each read is then assigned to the bin having the highest estimated probability of this read to belong to a given bin. A drawback of this abundance-based approach is to be only able to distinguish species whose abundance levels are considerably different.

Noting that most existing method for scaffold binning (such as SOM) require the user to manually and individually select bins, the authors of {\bf MaxBin} \citep{Wu2014} propose a fully automated scaffold binning model-based algorithm. Using expectation maximization over 4-mers frequencies as well as scaffold coverage levels, MaxBin can estimate the probability that a sequence belongs to a particular bin. Similarly to other model-based clustering algorithms, the results of  MaxBin are highly sensitive to parameter initialization, especially the number of bins. To circumvent this difficulty,  MaxBin can use results from FragGeneScan to identify single-copy marker genes and thus derive estimates for the number of bins.

\subsection{Multiple-sample binning}
\label{multiple-bin}

In response to the inability of abundance-based methods to separate species with similar abundances, a number of techniques have emerged that counter-balance it by considering multiple samples simultaneously. For example, algorithms proposed in {\bf MultiBin} \citep{Baran2012} start by pooling all reads from all samples and comparing them by pairwise alignment. The algorithm then performs $k$-medoid clustering on vectors representing the coverage in each sample.

More extensively, {\bf MultiMetaGenome} \citep{Albertsen2013} improves on this idea by considering the abundance difference in any of the samples in order to separate bins.  MultiMetaGenome applies a primary binning by clustering assembled genes by similarity, thus producing a non-redundant gene catalogue. Original reads are then mapped against these representative sequences and a normalized coverage for each sequence is computed. The representative sequences are further clustered into population genomes by plotting the two coverage estimates of all representative sequences against each other. In a similar way \citep{Nielsen2014} introduces a method for binning based on  clustering of co-abundant genes, where groups of genes are binned based on their abundance across multiple samples. Seed genes are picked randomly and bins (called co-abundance gene groups CAGs) are defined by correlation in terms of abundance. 

Recently, methods that combine information on both sequence abundance (as indicated by coverage) in multiple samples and sequence composition (or similarity) have  received a lot of attention. For example, {\bf Concoct} \citep{Alneberg2014} uses  Gaussian mixture models to cluster contigs based on both sequence composition and coverage across multiple samples. {\bf MetaBat} \citep{Kang2015} is a binning  method that uses a probabilistic distance based on the tetranucleotide frequency and coverage of contigs. This pairwise distance between contigs is subsequently used in a medoid clustering algorithm to consolidate similar contigs into bins.

\section{Diversity profiling and taxonomic assignment}
\label{sec:tax}

While OTU clustering and binning provide a global view of a metagenomic sample, they do not answer the question of which species constitute it. This is the goal of {\bf diversity profiling} and {\bf taxonomic assignment}. Diversity profiling aims to investigate the community structure of metagenomes, by providing abundances of different taxa. In the case of taxonomic assignment, we are interested to know to which taxon belongs each read (or assembled contig). Here we have a set of known reference  sequences $S=\{s_i\}$ and a set of class labels $L=\{l_i\}$ (species or any other taxonomic unit). Each sequence $s$ is labelled by its class. A taxonomic assignment algorithm uses $S$ in order to identify patterns in $S$ and thus be able to classify novel (eventually unknown) metagenomic sequences $M$ by providing a mapping $M \rightarrow L$. Notice that diversity profiling can be performed based on the results of taxonomic assignment, however dedicated methods aim at circumventing this step and rather to perform a global analysis for the reasons of computational efficiency.

Taxonomic assignment may take as input the result of OTU clustering or binning, or it can be done directly from sequencing data. Notice that many software methods provide a "two-in-one" solution integrating both clustering / binning and taxonomic assignment of the resulting clusters. As for OTU clustering and binning, taxonomic assignment methods are based either on sequence similarity provided by alignment or on genomic signatures in terms of oligonucleotide ($k$-mer) composition. Alignment-based methods combined with an algorithmic and data structure-driven solution (e.g. LCA) are very efficient for small-scale studies, such as amplicon sequencing. In particular, they provide state-of-the-art performance in terms of classification accuracy. Since these methods do not rely on machine learning, they will not be discussed in this review (see for review \citep{Bazinet2012}). See Table \ref{fig:tax-tools} for methods that we discuss in this review.

Even if they demonstrated state-of-art accuracy, alignment-based approaches are time-consuming and highly dependent on the presence of reference genomes in databases. Consequently, whole and complex metagenomic sample can hardly be analyzed using such methods. By contrast, machine learning alignment-free methods are based on pattern recognition, and should (theoretically) be more robust to noise or missing data. Furthermore, in terms of computation, once a model has been trained using reference data, it's execution on novel sequences is generally linear and thus much less time-consuming than alignment based methods. A recent study by Vervier and co-authors \citep{vervier2015large} that explores the possibility of a large-scale machine learning implementation for taxonomic assignment problem, has confirmed that compositional approaches achieve faster prediction times and consequently are appropriate for whole metagenome studies.

Taxonomic assignment relies on  reference databases in order to assign reads to taxonomic units. Consequently, computational methods fall into the category of supervised classification and learning, classification labels being provided by the taxa present in the reference database. However, there are several competing taxonomies that differ substantially \citep{DeSantis2006, Santamaria2012} and choosing among them is not necessarily obvious \citep{TangoMacha}. Moreover, for the unknown taxa, the inherent risk is to incorrectly assign their reads since they may be closely related to some of the reference sequences. Moreover, it has been long known that the assignment of short fragments (less than 1000bp) is a difficult task \citep{mchardy2007s}, consequently as for binning, a pre-assembly step is often performed \citep{Teeling, Mende}.

\begin{center}
\begin{table}[htbp]
\includegraphics[width=\textwidth]{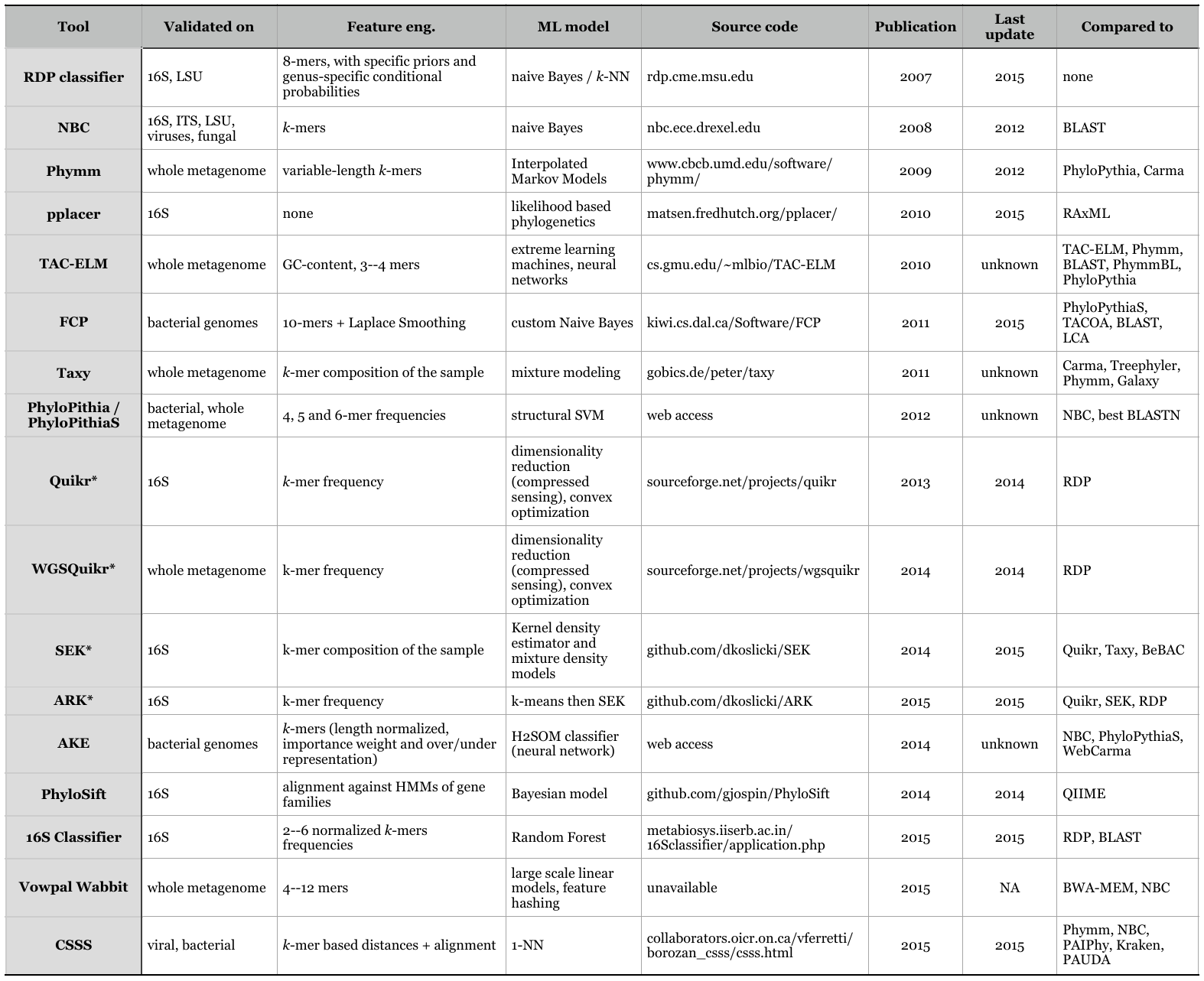}
\caption{\label{fig:tax-tools} Characteristics of tools for diversity profiling (signalled by $^*$) and taxonomic assignment. Shown are types of data on which the method has been {\bf validated}, how features have been extracted (column {\bf Feature eng.}), the machine learning algorithm (column {\bf ML method}), link to the {\bf Source code}, year of {\bf publication} as well as the last source code update (column {\bf Last update}), and finally tools against which the performance has been evaluated (column {\bf Compared to}).}
\end{table}
\end{center}
\vspace*{-3em}

\subsection{Na{\"i}ve Bayes and Bayesian methods}
\label{sec:tax-nb}

The well-known ribosomal database project {\bf RDP classifier} \citep{Wang2007,cole.ea:2009} relies on a reference sequence database which contains relevant species, then assigns a class label (e.g. species, genus, etc.) to each read by the na{\"i}ve Bayesian algorithm based on $k$-mer occurrence information. Similarly, the popular {\bf NBC} \citep{Rosen2011} tool and {\bf FCP} classifier \cite{Parks2011} implement a na{\"i}ve Bayesian framework based on $k$-mer counts. Notice however, that RDP classifier is meant for 16S data, while NBC was developed for whole genome sequencing projects.

While based on Bayes’ rules, Na{\"i}ve Bayesian classifiers, such as learnt by RDP and NBC, are not bayesian in the formal sense, as they do not derive probability distributions over posteriors. By contrast, {\bf pplacer} \citep{Matsen2010} offers a full probabilistic and Bayesian framework to locate a query sequence in a reference phylogeny. From this localisation, a taxon identifier can be assigned to the query sequence.  pplacer is used as a component of the popular {\bf PhyloSift} \citep{Darling2014} suite, that uses a reference database of gene families, aligns all of the input reads against this database and then calculates the posterior probability that the read diverged from particular branches of the reference tree via direct integration.

\subsection{Model-based methods}
\label{sec:tax-model}

{\bf Phymm} \citep{Brady2009} uses Interpolated Markov Models to learn variable-length oligonucleotides typical of a given phylogenetic grouping. In a similar fashion, {\bf PhyloPithia} and {\bf PhyloPithiaS} \citep{McHardy2007,Patil2012} learn Support Vector Machine classifiers based on $k$-mer frequencies associated with different taxonomic groups. This classifier is then used to assign reads from a new metagenomic sample to pre-existing taxa.

The training step for such approaches can be quite slow, which gave rise to software tools that employ different optimization techniques. These approaches target the problem of diversity profiling, for which establishing a global model is particularly well-suited. Among these methods are convex optimization strategies such as {\bf Quikr} \citep{Koslicki2013} (and its whole genome version {\bf WGSQuikr} \citep{ WGSQuikr2014}) and {\bf Taxy}, a mixture modeling approach of the overall oligonucleotide distribution of a metagenomic dataset \citep{Meinicke2011}. {\bf SEK} \citep{Chatterjee2014} aims to overcome the sensitivity of Quikr to the differences between length of sequences that are present in the database and read length, by proposing a method based on kernel density estimators and mixture density models, that enables the estimation of taxonomic units' frequencies by solving an under-determined system of linear equations. {\bf ARK} \citep{koslicki2015ark} is another method for diversity profiling that aims to improve the precision of convex optimisations approaches by aggregating the reads from a sample with an unsupervised machine learning approach prior to the estimation phase. 

Other classical machine-learning approaches, such as Random Forests and neural networks have been advanced to perform taxonomic assignment. For example, {\bf 16S Classifier} \citep{Chaudhary2015} uses Random Forests for the taxonomic classification of short 16S rRNA and the Greengenes taxonomy for assignment, while {\bf TAC-ELM} \citep{rasheed2012metagenomic} is a composition-based method (oligonucleotides and GC content) that uses a neural network-based model. Moreover,  Vervier and co-authors have demonstrated that modern large scale linear classifiers with feature hashing such as {\bf Vowpal Wabbit} \citep{vervier2015large} can be efficiently used.

The recent {\bf CSSS} method \citep{Borozan2015} exploits the complementary nature of alignment-based and alignment-free similarity measures and applies the nearest neighbor clustering. In particular it aims at being able to assign taxonomic ranks to both bacterial and viral communities. However, having to compute four different similarity measures, impacts the scalability.

\section{Comparative Metagenomics}
\label{sec:comparative}

The questions that we have addressed so far concern the association of DNA fragments from a given sample with a class (either an OTU, a bin or a taxon). On the other spectrum of machine learning applications for metagenomics lie methods that try to assign labels to whole samples, by analyzing features derived from all DNA fragments that compose it. Such classification aims for example at analyzing phenotypes based on metagenomic fragments and has wide applications in biomedical settings.

Following the distinction between binning and taxonomic assignment, we first review unsupervised methods that aim at clustering / comparing samples that are not labeled then we review supervised methods that can classify metagenomic samples into pre-defined classes. See Table \ref{fig:comparative-tools} for a summary.

\begin{center}
\begin{table}[htbp]
\includegraphics[width=\textwidth]{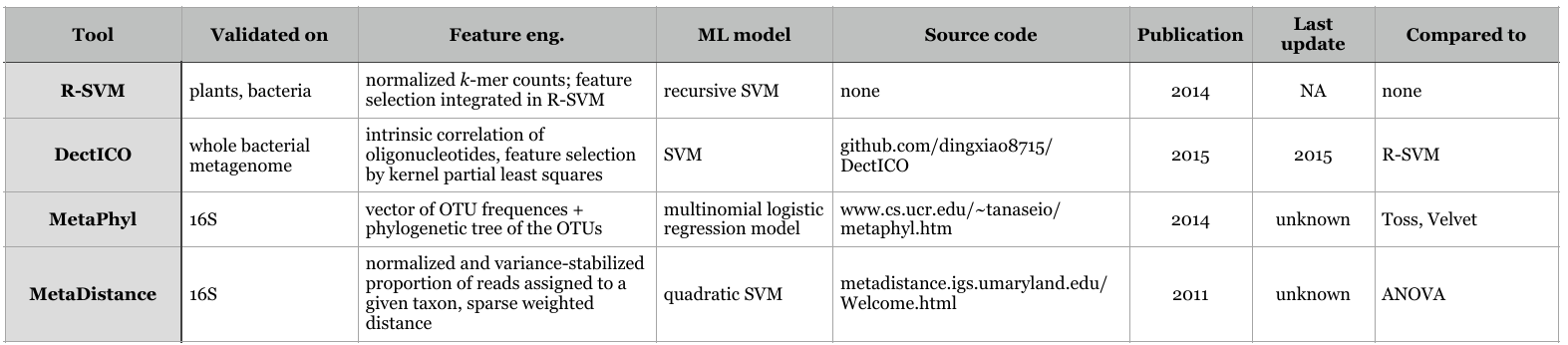}
\caption{\label{fig:comparative-tools}Characteristics of tools for comparative metagenomics. Shown are types of data on which the method has been {\bf validated}, how features have been extracted (column {\bf Feature eng.}), the machine learning algorithm (column {\bf ML method}), link to the {\bf Source code}, year of {\bf publication} as well as the last source code update (column {\bf Last update}), and finally tools against which the performance has been evaluated (column {\bf Compared to}).}
\end{table}
\end{center}
\vspace*{-3em}

\subsection{Unsupervised methods}
\label{sec:comparative-unsupervised}

One of the first approaches to assess the distance between metagenomic samples is known as the  {\bf Directed Homogeneity test} \citep{Mitra2009} and is implemented in the MEGAN toolbox. This test employs Pearson's chi-square test over all sub-nodes of a taxa, and accounts for multiple testing correction.  Although this test can indicate how significant is the difference between two samples, it is not usable in a distance-based clustering approach. 

Jiang and co-authors explore different sequence signature methods (14 $k$-mer based distances) and UPGMA clustering for the problem of clustering metagenomic samples as well as for recovering environmental gradients that affect microbial samples \citep{jiang2012comparison}. The authors concluded  that $k$-mer based descriptors are well-suited for this task and have singled out the $d_2^S$  dissimilarity measure, (corresponding to the uncentered correlation between the number of counts of $k$-mers for two sequences normalized by standard deviation) that outperformed others for both application scenarios. This measure has been recently employed in the DectICO method  \citep{ding2015} (see section \ref{sec:comparative-supervised}).

The {\bf DSM Framework} of \citet{Seth2014} aims at computing a dissimilarity measure between samples based on their $k$-mer frequencies. Such a measure can be used for initial exploratory analysis of a large set of metagenomic samples. The authors propose to compare samples based on the Jaccard dissimilarity between normalized relative frequencies of informative $k$-mers. By using an efficient distributed implementation, informative $k$-mers can be identified \emph{over all values of $k$ simultaneously}. Although this is an unsupervised method in the sense that samples are \emph{a priori} not labeled, the authors demonstrate that their method can discriminate samples more efficiently than supervised methods based on protein families. In our opinion, such feature estimation and feature selection can be used for other machine learning approaches aiming at sample classification.

\subsection{Supervised methods}
\label{sec:comparative-supervised}

The unsupervised approaches are powerful for exploratory data analysis, but are not adequate when samples are known to belong to different groups or phenotypes. In this setup, supervised machine learning aims at inferring the relationship between the sample's content and the label of interest. 

To the best of our knowledge, the first attempt at supervised learning for comparative metagenomics was performed by  \citet{yang2006} to distinguish between soil and sediment samples obtained via amplicon sequencing. They applied both Random Forest and $k$-nearest neighbors algorithms over all identified hyper-variable regions. Specifically for metagenomics samples, the {\bf MetaDistance} \citep{Liu2011} method was devised for multi-class classification of amplicon sequenced metagenomic datasets. MetaDistance represents samples as a normalized and variance-stabilized proportion of reads assigned to a given taxon. This matrix is then used to learn a sparse weighted distance by using a quadratic SVM formulation that maximises intra-class distance and minimized inter-class distance. This sparse distance encodes feature selection and is then used in a $k$-NN algorithm. 

Owing to the observation that OTUs can be organized as a tree, \citet{Tanaseichuk2014} proposed a multi-class metagenomic classifier called {\bf MetaPhyl} that accounts for OTUs relationships. Each sample is represented by a vector of OTU frequencies, and a multinomial logistic regression model with a tree penalty is then trained. The authors report greatly improved performance over MetaDistance and over generic classifiers (logistic regression and Random Forest).

Another approach specifically devised for metagenomic sample classification is the {\bf R-SVM} of \citet{cui2013} where the authors generalized recursive SVMs to perform both feature selection and discrimination of human whole metagenome samples for both control and inflammatory bowel disease patients. 

More recently, this approach has been generalized in the  {\bf DectICO} tool \citep{ding2015}, which combines (i) kernel partial least squares as a strategy for feature selection,  (ii) \emph{intrinsic correlation of oligonucleotides} (ICO) that generalizes the $k$-mer frequencies to describe samples and (iii) SVM as a learning algorithm. The authors claim that DectICO method outperforms other approaches when long $k$-mers are considered. 

\section{Gene prediction}
\label{sec:functions}

Another broad question that metagenomics tries to answer is the {\bf functional analysis} for metagenomic studies, that is determining what are the functional and metabolic repertoires of a given community, what are the differences between communities and what are the members that enable them to exert different effects. Ideally, a functional profiling of a community would establish a complete geneset for each  species present in the sample and perform further functional annotation and network analysis. This is still not possible.

Considering a microbial community as a whole (and thus ignoring the exact assignment of a gene to a specific species) gives rise to community network analysis. It has even been argued that a functional view should better characterize a microbial community than its taxonomic composition since different organisms are able to perform similar biological functions, giving rise to the concept of community-realized functions \citep{Liu2012, Fuhrman2009, Zhou2010}. To the best of our knowledge, neither functional annotation nor network reconstruction have been tackled by machine learning approaches. Consequently, this section is dedicated to gene prediction in metagenomic samples, see Table \ref{fig:gene-tools} for an overview. 

Gene prediction methods attempt to identify patterns within DNA sequences that correspond to those recognized by transcription and translation machinery. Typically, gene prediction tools predict the start and stop sites of protein-coding genes and produce \emph{in silico} translations of these genes. 

Standard tools for gene finding were designed to work on single genome sequencing data (e.g. Glimmer \citep{Delcher1999}). Typically, gene prediction is based on Hidden Markov models that are trained on the gene structure of known organisms similar to the one that is being studied or alternatively on generalized models of prokaryotic or eukaryotic genes. Often, genes present in a metagenomic sample come from different organisms, this is why such an approach can not produce predictions with high confidence. Indeed, true positive rates of gene identification in metagenomic samples are at best around 70\% of true positives for reads,  missing a significant subset of genes \citep{DeFilippo2012, Trimble2012}. Since the accuracy in complete genomes is very high ($>95\%$), an assembly step can greatly improve gene prediction for metagenomic samples and some methods explicitly aim at longer sequences.

It has also to be noted that most of the proposed computational methods target bacterial communities. Thus, gene prediction for mixed metagenomic datasets remains a largely unresolved question.

\begin{center}
\begin{table}[htbp]
\includegraphics[width=\textwidth]{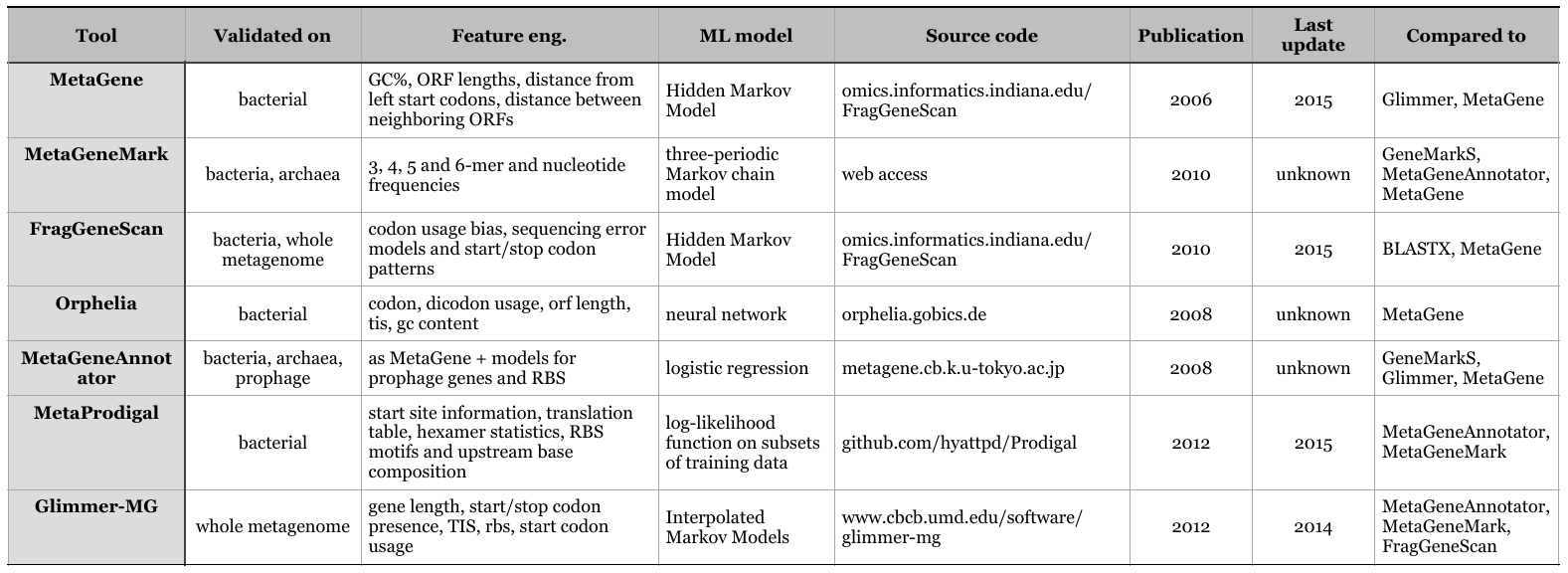}
\caption{\label{fig:gene-tools} Characteristics of gene detection tools. Shown are types of data on which the method has been {\bf validated}, how features have been extracted (column {\bf Feature eng.}), the machine learning algorithm (column {\bf ML method}), link to the {\bf Source code}, year of {\bf publication} as well as the last source code update (column {\bf Last update}), and finally tools against which the performance has been evaluated (column {\bf Compared to}).}
\end{table}
\end{center}
\vspace*{-3em}
 
\subsection{Statistical methods}
\label{sec:functions-stat}

One of the first gene prediction methods for metagenomic datasets, {\bf MetaGene} \citep{Noguchi2006}, uses logistic regression models whose features are the GC content and the di-codon frequencies in order to distinguish between gene-coding and non-gene coding ORFs. {\bf MetaGeneAnnotator} \citep{Noguchi2008} builds on the MetaGene approach by integrating more thorough statistical models for bacterial, archaeal and prophage genes and incorporating species-specific patterns of ribosome binding sites in order to increase the confidence of the translation starts site prediction. 

{\bf MetaProdigal} \citep{Hyatt2012} is a method that improves on the Prodigal software \cite{Hyatt2010} and is based on the same idea of gene prediction. The coding regions are distinguished from the background by using a log-likelihood function that incorporates information such as translation table, hexamer statistics, ribosomal binding site motifs and upstream base composition. The main difference with Prodigal resides in the fact that training data is pre-clustered using a complete-linkage clustering and the log-likelihood function is defined on these clusters. 

\subsection{Model-based methods}
\label{sec:functions-markov}

{\bf MetaGeneMark} \citep{Zhu2010} is built on the same principle as the popular HMM-based gene prediction tool GeneMark, but adapts it to process short DNA fragments. First, training genomes are used to estimate a polynomial and logistic approximations of oligonucleotide frequencies as a function of the GC-content. These estimates are then used to set the parameters of the MetaGeneMark gene finder HMM. {\bf FragGeneScan} \citep{Rho2010} is another HMM-based approach that integrates codon usage bias, start/stop codon patterns as well as an explicit modeling of sequencing errors; it  seems to outperform other solutions and is currently used by the EBI Metagenomics pipeline.
 
{\bf Glimmer-MG} \citep{Kelley2011} relies on Interpolated Markov Models to identify the coding regions and distinguish them from noncoding DNA. First, binning is performed using SCIMM and the IMM models are trained within each cluster for gene predictions, which makes this method quite computationally expensive. 

{\bf Orphelia} \citep{Hoff2008, Hoff2009} is a two-stage machine learning approach. First, linear discriminants for monocodon usage, dicodon usage and translation initiation sites are used to extract features from  sequences. Second, ORFs are classified as being protein-coding or not by a neural network that combines these features with ORF length and GC-content. The neural network is trained on random sub-sequences of a given length of genomes in the reference database.

\vspace*{1em}
In addition to the novel computational methods presented in this review a number of integrated pipelines has been developed such as the EBI Metagenomics Webserver \citep{hunter2014ebi} based on FragGeneScan, RAMMCAP \citep{li2009analysis} based on MetaGene and SmashCommunity based on GeneMark and MetaGene \citep{arumugam2010smashcommunity}.

\section{Discussion}

Applications of machine learning in metagenomics have penetrated all major questions asked in this field of research. Moreover, this idiom of data analysis has been a major driver in all areas of metagenomics: OTU clustering, binning, taxonomic assignment, comparative metagenomics and functional analysis. In turn, these biological applications have triggered many original developments in machine learning: in the way data are represented, how relevant features are selected and how prior knowledge is introduced in the algorithm. 

Looking at the tables that recapitulate the examined tools, we can notice a clear progression towards the sophistication of machine learning approaches. Even better, today certain metagenomic questions can be considered to be essentially solved, in the sense that they can be answered in a computationally efficient way and tend to produce high quality results. The most noteworthy example is the taxonomic assignment for bacterial communities in the case of amplicon sequencing (for the cases where the resolution provided by reference taxonomies is considered to be sufficient for the study). 

Despite the spectacular technological and methodological advances, the science of metagenomics still has a large number of questions for which no definitive answer has been provided. Here we enumerate some of these questions, which in our opinion constitute the major avenues for current and future research in metagenomics. These questions pertain to both biological and computational aspects of metagenomics. 

\begin{itemize}
\item \emph{Novel species}. Dealing with unknown species complicates both computation and interpretation. For example, OTU clustering and binning can be efficiently performed; however, interpretation of the resulting groups is difficult. At the same time, taxonomic assignment is unable to assign sequences from such species if no sufficiently close taxa exist in the reference databases.

\item \emph{Whole metagenomic samples}. As we have seen, many computational approaches have been validated on bacterial data. Being able to efficiently and accurately deal with samples containing, for example, host traces, archaea, viruses, fungi, protists and small algae, is still an open question.

\item \emph{Functional analysis}. Detection of protein-coding genes and consequently their downstream analysis remain challenging. In particular this involves developing methods able to efficiently analyse heterogeneous and fragmented data in the presence of sequencing errors.

\item \emph{Big Data challenge}. Machine learning is still seldom deployed in the ``Big Data'' fashion. Many authors remark that current methods cannot be applied to datasets with sizes over a Terabyte. This constitutes a bottle-neck for large-scale environmental studies. An effort in porting existing methods to modern Big Data machine learning frameworks (e.g., Spark etc.) or in developing novel ones has to be undertaken.
\end{itemize}

NGS has catalysed the research in metagenomics, and paved the way for scientists to build fundamental knowledge on various communities over the past decade. The technical ability to routinely generate large datasets of sequences from extremely variable communities has enabled the study of different environments in a less biased way. Part of the solution to the challenges identified above lies in the technical advances for data generation, especially longer reads. The field is finally mature enough to think ``beyond the metagenomics'' and advance our understanding of microbial communities and different environments by including such aspects as, for example, host-pathogen interactions, quorum sensing and cell-to-cell interactions. This would require collecting additional experimental measurements and metadata, and in turn call for yet another generation of computational methods for what we can call \emph{integrative metagenomics}.

\paragraph{Acknowledgements}
The authors would like to thank the anonymous reviewers for their helpful comments.
This work was partly supported by R\'egion Aquitaine grant 2014.

\bibliography{metagenomics-ml}

\end{document}